\documentclass[twoside,12pt]{article}
\usepackage{epsfig}
\usepackage{hyperref}
\usepackage{epsfig}
\usepackage{graphicx}
\usepackage{amssymb,amsmath}

\newcommand{\be}{\begin{equation}}
\newcommand{\ee}{\end{equation}}
\newcommand{\bea}{\begin{eqnarray}}
\newcommand{\eea}{\end{eqnarray}}

\begin{document}

\title{Coarse graining Nuclear Interactions~\footnote{Presented by
    R.N.P. in  {\it From Quarks and Gluons to Hadrons and Nuclei}
33rd Course International School of Nuclear Physics, Erice, Sicily, 16-24 September 2011.}}
\author{\underline{R. Navarro P\'erez}, J.E. Amaro and E. Ruiz Arriola
  \\ Departamento de F\'{\i}sica At\'{o}mica, Molecular y Nuclear,
  \\ Universidad de Granada, E-18071 Granada, Spain.}  \maketitle
\begin{abstract} 
We consider a coarse graining of NN interactions in coordinate space
very much in the spirit of the well known $V_{\rm low k}$ approach. To
this end we sample the interaction at about the minimal de Broglie
wavelength probed by NN scattering below pion production
threshold. This amounts to provide a simple delta-shell potential plus
the standard OPE potential above 2 fm.  The possible simplifications
in the Nuclear many body problem are discussed.
\end{abstract}
\section{Introduction}
The NN interaction provides a basic building block of atomic nuclei.
A milestone in the development of the field was achieved when the
Nijmegen group generated a fit via a partial wave analysis (PWA) to a
set of about 4000 NN scattering data with $\chi^2 /{\rm dof} \lesssim
1 $~\cite{Stoks:1993tb} after charge dependence (CD) effects were
incorporated and discarding about further 1000 of $3\sigma$
inconsistent data.  The analysis was carried out using an energy
dependent potential for which nuclear structure calculations become
hard to formulate. Thus, energy independent {\it high quality}
potentials were produced with almost identical $\chi^2/{\rm
  dof}$~\cite{Stoks:1994wp,Wiringa:1994wb}.  Among them, the AV18
potential is directly useful for {\it ab initio} Monte Carlo
calculations up to $A=10$~\cite{Pieper:2001mp}. While all these
potentials differ in their form, in the last years it has been
realized that if CM momenta above $\Lambda= 400 {\rm MeV}$ are
explicitly integrated out, the remaining effective interaction has
appealing features.  The so-called $V_{\rm low k}$
potentials~\cite{Bogner:2003wn} exhibit an astonishing degree of
universality, produce a rather smooth interaction and weaken the
strength of the interaction so that Hartree-Fock calculations may be
reliable starting points for nuclear structure calculations. In the
present talk we adress a suitable formulation of the problem in
configuration space.

\begin{figure}[tb]
\begin{center}
\begin{minipage}[t]{6 cm}
\epsfig{file=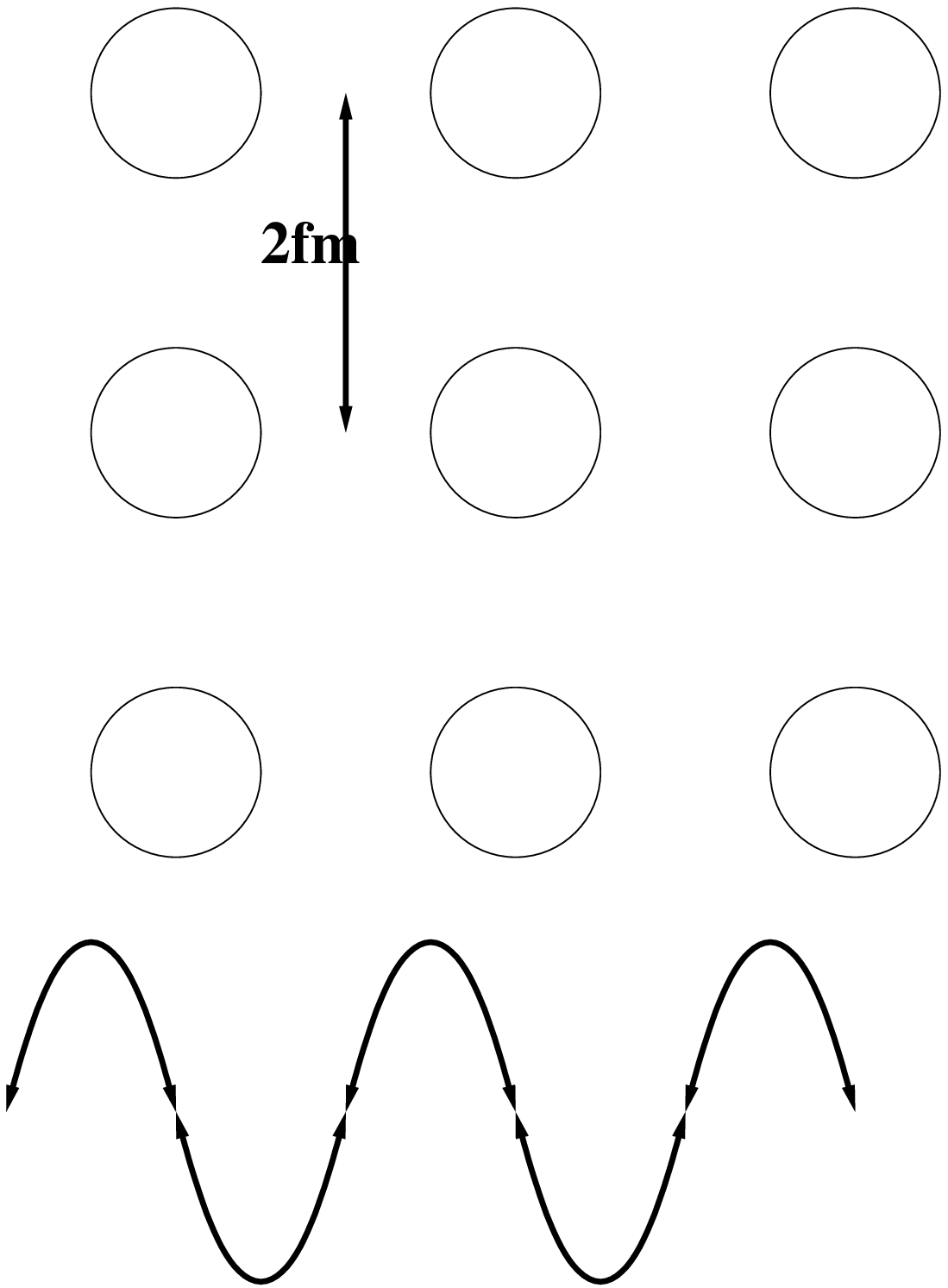,scale=0.3}
\end{minipage}
\begin{minipage}[t]{8 cm}
\epsfig{file=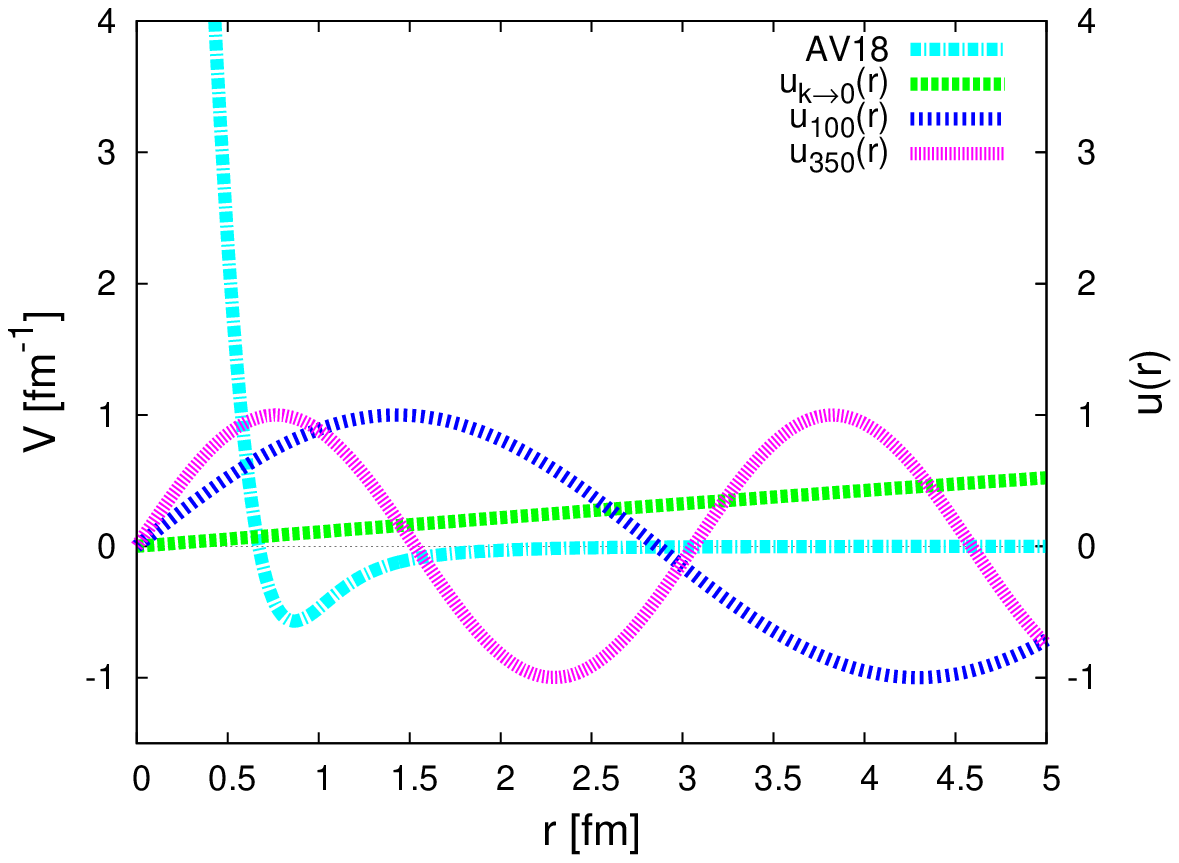,scale=0.6}
\end{minipage}
\begin{minipage}[t]{16.5 cm}
\caption{Left panel: Cartoon of a nucleus, displaying the size of the
  nucleons as compared to the typical distance to nearest
  neighbors and the shortest wave-lengths wave functions. 
 Right panel: The AV18-potential in the $^1S_0$ channel. Superposed are 
the eigen wavefunctions at zero energy, $E_{\rm LAB}=100$ MeV and 
$E_{\rm LAB}=350$ MeV.\label{fig:nucleus}}
\end{minipage}
\end{center}
\end{figure}

Regarless of these successes it is to date unclear what is the impact
of NN uncertainties on finite nuclei due to our ignorance on short
distances. Relevant length scales are a) The mean interparticle
separation distance $d = 1.8 {\rm fm}$ as obtained from Nuclear matter
saturation density $\rho_0 = 1/d^3 = 0.17 {\rm fm}^{-3}$, b) The Fermi
momentum $k_F = (3/2)^{\frac13}/d \sim 250 {\rm MeV}$ which gives a
wavelength of about $\hbar/k_F = 0.8 {\rm fm}$, c) Minimal relative CM
de Broglie wavelengh corresponding to the pion production threshold
$\lambda=\hbar / \sqrt{M_N m_\pi} \sim 0.5 {\rm fm} $ and d) The pion
Compton wavelength $1/m_\pi = 1.4 {\rm fm}$.  The situation is
presented pictorially in Fig.~\ref{fig:nucleus}
suggesting that for the  description of the ground state in light nuclei 
both the short distance core and the role of explicit pions become
marginal. This was recognized long ago~\cite{Afnan:1968zj} where the
bulk of $^3{\rm He}$ and $^4{\rm He}$ could be described with a
pionless and soft-core potential which just reproduced the S-wave
phase shifts up to $E_{\rm LAB} \le 100 {\rm MeV}$. Actually, we
expect this feature to hold for light nuclei.

\section{The delta-shells (DS) potential}

\begin{figure}[tb]
\begin{center}
\begin{minipage}[t]{7 cm}
\epsfig{file=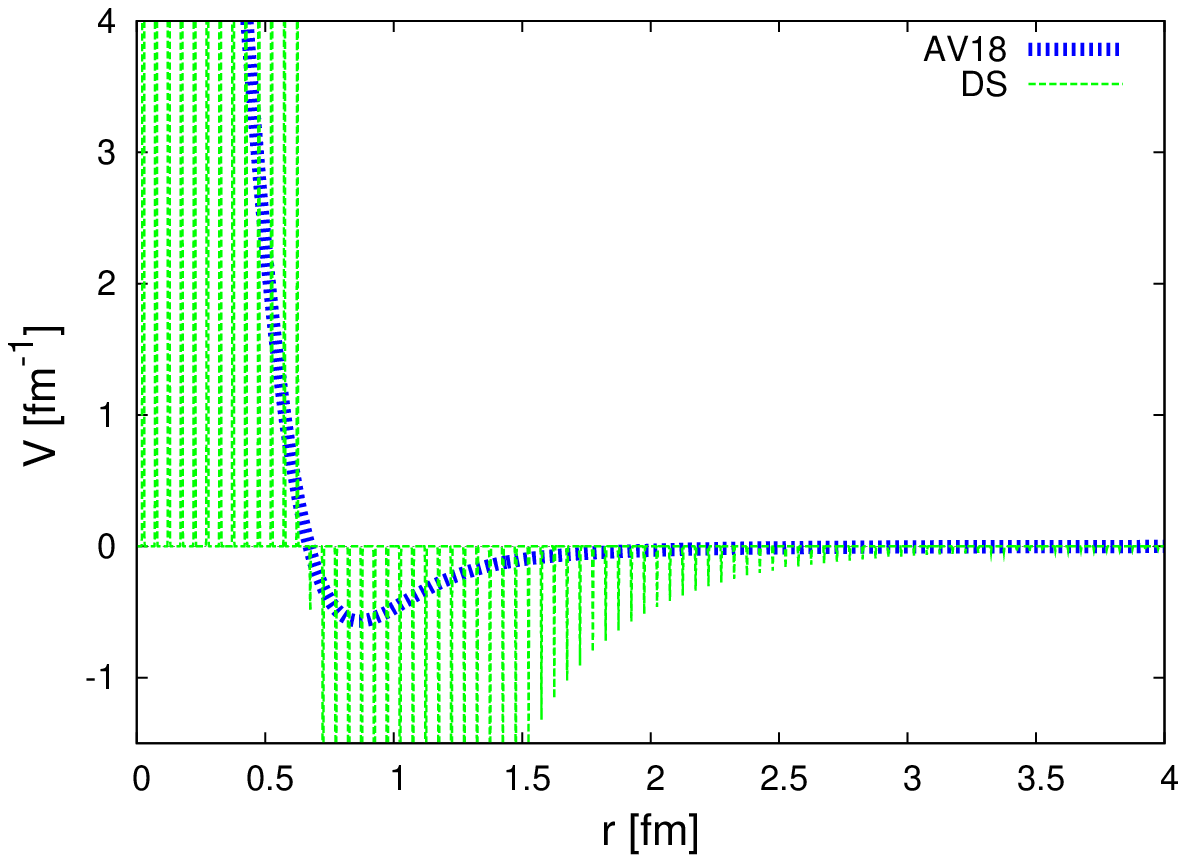,scale=0.5}
\end{minipage}
\begin{minipage}[t]{7 cm}
\epsfig{file=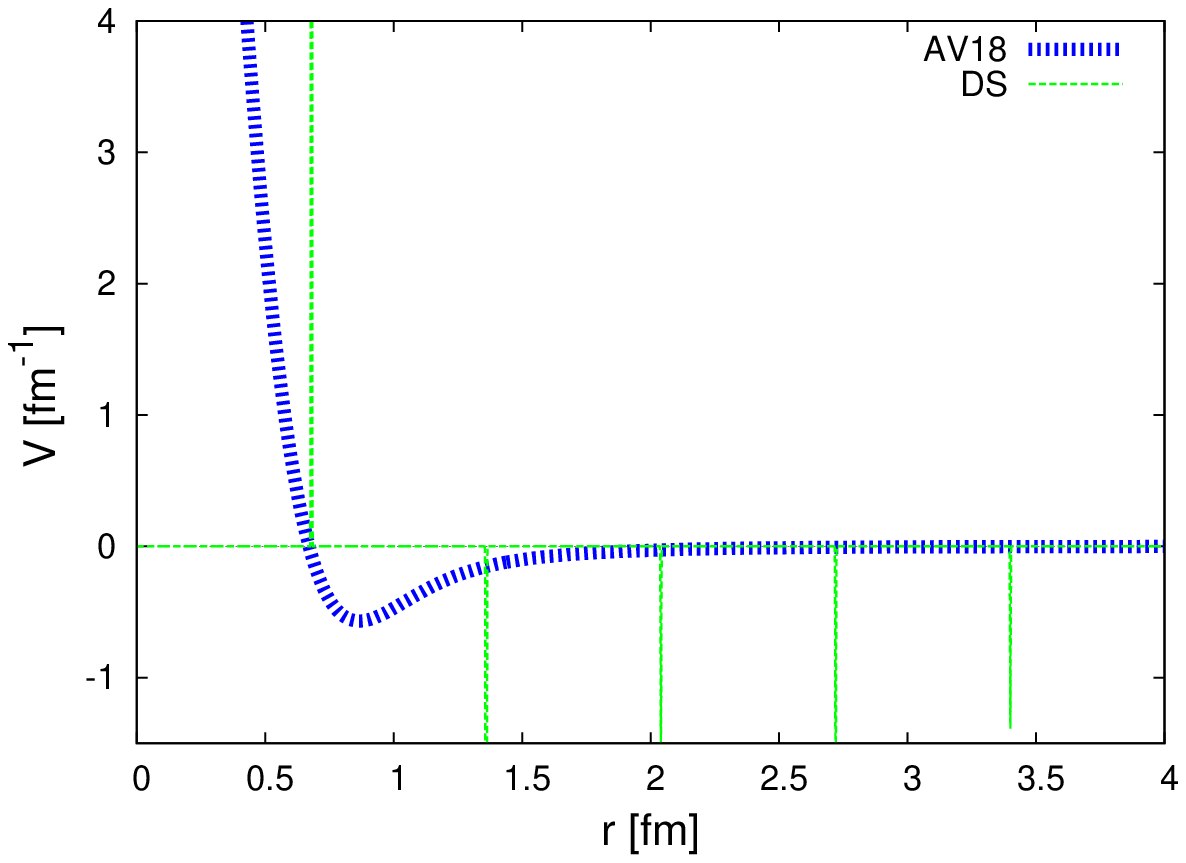,scale=0.5}
\end{minipage}
\begin{minipage}[t]{7 cm}
\epsfig{file=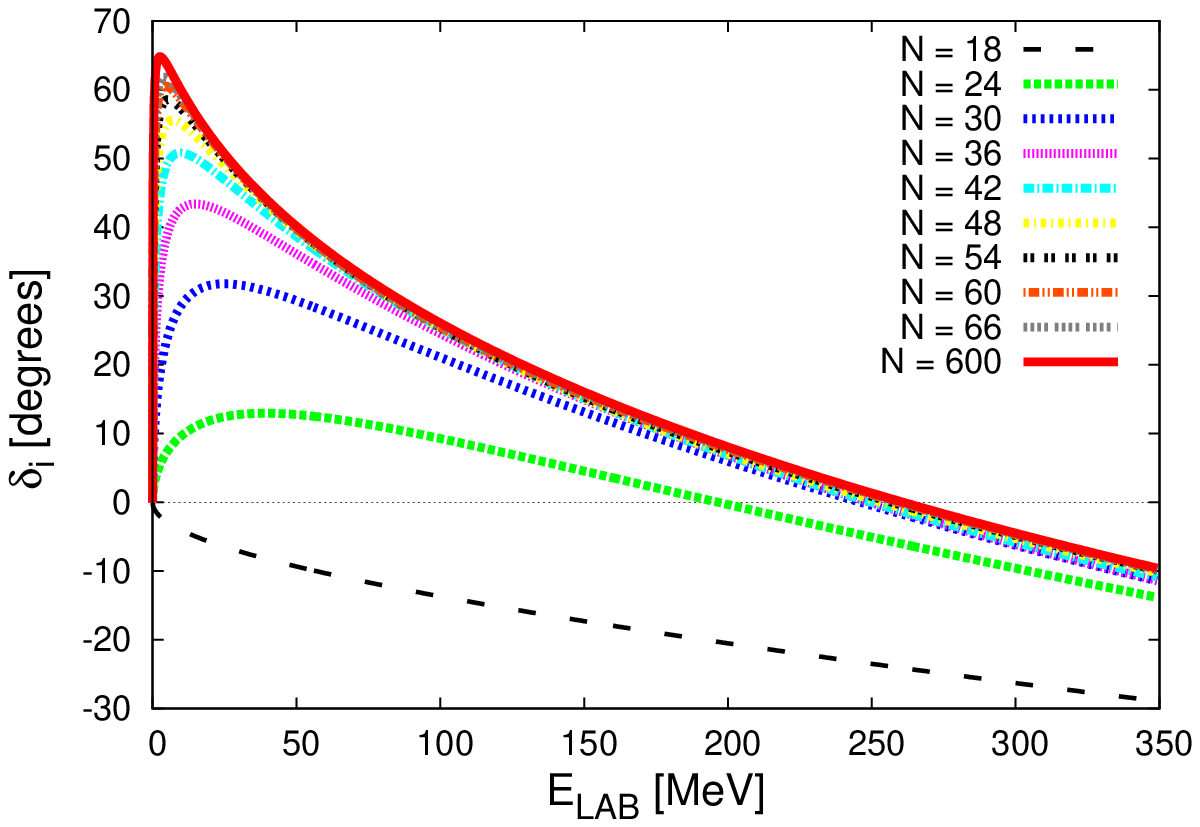,scale=0.5}
\end{minipage}
\begin{minipage}[t]{7 cm}
\epsfig{file=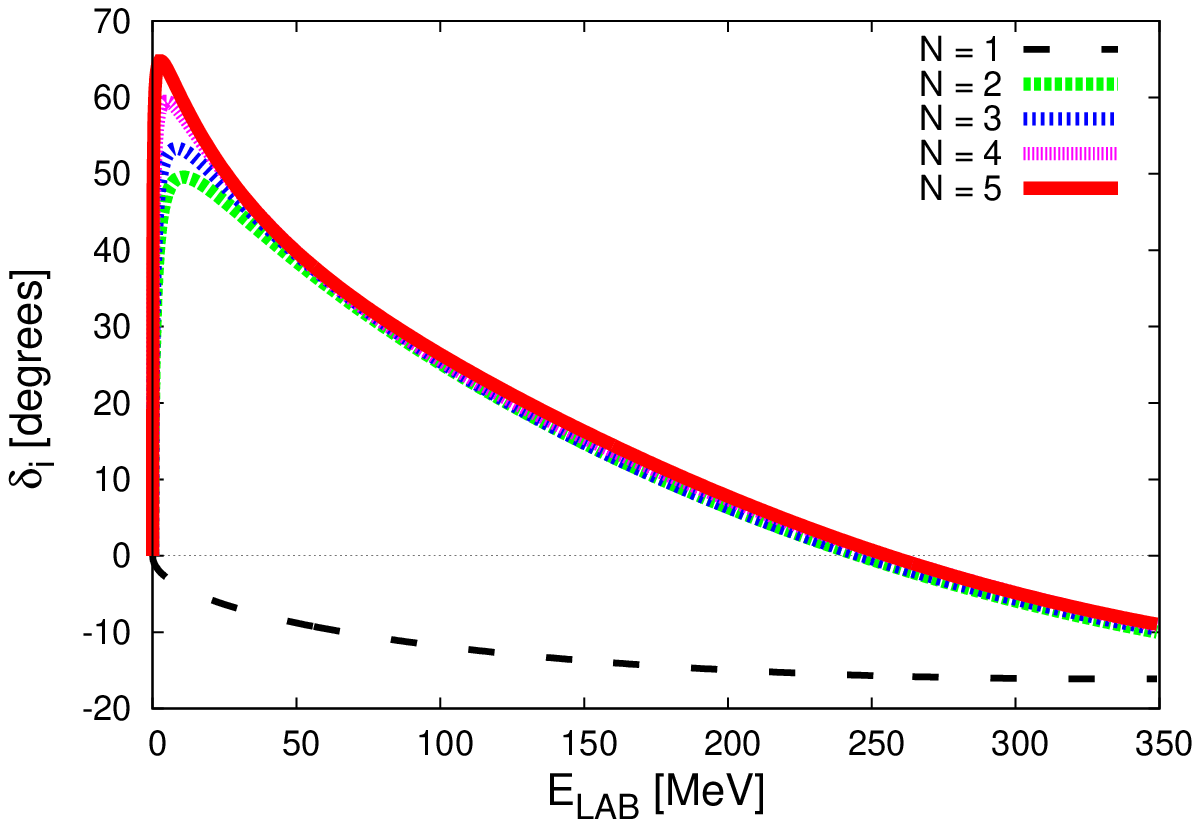,scale=0.5}
\end{minipage}
\begin{minipage}[t]{16.5 cm}
\caption{Upper panel: The realistic AV18 potential ($^1S_0$-channel)
  and its delta-shell representation (left) and the coarse grained
  potential (right). Lower panel: Accumulated phase shifts for both
  cases as a function of the LAB energy (in MeV) indicating the number
  of grid points. Both resulting phase-shifts 
  coincide.\label{delta-shell}}
\end{minipage}
\end{center}
\end{figure}

The basic observation was made long ago by Aviles~\cite{Aviles:1973ee}
and recently rediscovered in the context of renormalization of chiral
forces~\cite{Entem:2007jg}. If the two-particle CM wave numbers are
limited to a range $\Delta k$ only gross information can be determined
in an interval $\Delta r$, (see e.g. Fig.~\ref{fig:nucleus}) with
$\Delta r \Delta k \sim 1 $. Thus, for $\Delta k = \Lambda \sim 400
{\rm MeV}$ we have $\Delta r_{\rm min} \sim 0.5 {\rm fm}$. This
uncertainty suggests that for a limited energy range the potential
only needs to be known in a limited number of points. With this in
mind we consider a neutron-proton (np) potential as a sum of $\delta$
functions
\begin{equation}
\label{eq:DeltaShellPotential}
V(r) = \sum_{i=1}^{n}{ \frac{\lambda_i}{2 \mu} \delta (r - r_i)}  \, ,  
\end{equation}
where $\mu = M_N/2$ is the reduced np mass of the system, the $\lambda_i$
coefficients are strength parameters and $r_i$ are the concentration
radii. In that case we may determine the s-wave as
\begin{eqnarray}
u(r) &=& \sin ( k r + \delta_{i+1/2}) \, , \quad r_i \le r \le r_{i+1} \, ,  
\label{eq:vf_u} 
\end{eqnarray} 
where $\delta_{i+1/2}(k)$ is the accumulated phase shift at the mid-point
$r_{i+1/2}$ and $\delta(k) \equiv \delta_{N+\frac12}(k)$. 
Matching the discontinuity of log-derivatives at
$r=r_i$, we simply get
\begin{eqnarray}
k \cot ( k r_i + \delta_{i+1/2} ) - k \cot ( k r_i + \delta_{i-1/2} )
&=& \Delta r U(r_i) \, ,  
\label{eq:vf_dis} 
\end{eqnarray} 
where $U(r) = 2 \mu V(r) = M V(r)$ is the reduced potential.  The
regular solution at the origin requires $ \delta_{\frac12} (k)=0 $. If
we take the limit $\Delta r \to 0 $ we can define $\delta ( k , r_i )
= \delta_i (k) $, to get
\begin{equation}
\frac{ d \delta (k,R) }{dR} = -\frac1k U(R) \sin^2 (k R+ \delta(k,R))
+ {\cal O} ( \Delta r^2 )
\, . 
\label{eq:vf}
\end{equation}
which is the variable phase equation~\cite{Calogero:1965} up to finite
grid corrections and can be interpreted as the change in the
accumulated phase when a truncated potential of the parametric form $U
(r) \theta (R-r)$ is steadily switched on as a function of the
variable $R$. This equation and its generalization to coupled channels
has extensively been used to treat the renormalization problem in NN
scattering in
Refs.~\cite{PavonValderrama:2003np,PavonValderrama:2004nb,PavonValderrama:2007nu}. The
low energy expansion of the discrete variable phase equations was used
in Ref.~\cite{PavonValderrama:2005ku} to determine threshold
parameters in all partial waves. The relation to the well-known
Nyquist theorem of sampling a signal with a given bandwidth has been
discussed in Ref.~\cite{Entem:2007jg}.  Of course, this DS
approximation to the potential can be most immediately used as a
numerical method to solve the scattering problem, which would become
exact for $\Delta r \to 0$, if we take the weights {\it given} by the
potential $\lambda_i = U(r_i) \Delta r_i $. As an illustration we show
in Fig.~\ref{delta-shell} the phase-shifts obtained for the
$^1S_0$-AV18 potential for several values of $\Delta r$. Convergence
to the phases to four significant figures is achieved for $\Delta r
=0.01 {\rm fm}$.  The equidistant discretization corresponds to the
trapezoidal rule and one could improve by a more sophisticated method,
a relevant issue when the interaction is known {\it a priori}.

\section{Coarse grained local potentials}

Another, more fruitful and economical, perspective already pursued by
Aviles corresponds to consider the weights themselves, $\lambda_i$, as
fitting parameters to the phase-shifts, since anyhow the potential at
short distances is unknown and will be determined from the data.

If we take just one delta-shell in S-wave we may determine both the
point $r=r_c$ and its corresponding strength, $\lambda_c$ from the
scattering length $\alpha_0$ and the effective range $r_0$, defined 
from 
\begin{eqnarray}
k \cot \delta_0 (k) = - \frac{1}{\alpha_0}+ \frac12 r_0 k^2 + \dots 
\label{eq:ere}
\end{eqnarray}
 For
instance, for the $^1S_0$ case one has $\alpha_{0,^1S_0}= -23.74 {\rm
  fm}$ and $r_{0,^1S_0}= 2.77 {\rm fm}$ which yields $\lambda_{^1S_0}
= -0.4626 $ and $ r_{c,^1S_0} = 1.99 {\rm fm}$ whereas for the $^3S_1$
channel one gets $\alpha_{0,^3S_1}= 5.42 {\rm fm}$ and $r_{^3S_1}=
1.75 {\rm fm}$ giving $\lambda_{^3S_1} = -0.911 $ and $ r_{c,^3S_1} =
1.53 {\rm fm}$.  The corresponding phase shifts are reproduced to
about $E_{\rm LAB} \lesssim 50 {\rm MeV}$. One can improve on this by
including more delta-shells. A good fit to the PWA is obtained for
($\lambda_i$ in ${\rm fm}^{-1}$ and $r_i$ in ${\rm fm}$)
\begin{eqnarray}
(\lambda_i, r_i) &=& ( -0.568, 1.56) \, , (-0.023, 3.47) \, ,\quad  ^1S_0 \, ,\nonumber \\ 
(\lambda_i, r_i) &=&  (-0.951, 1.35) \, , (-0.052, 2.60) \, , \quad ^3S_1 \, .
\label{eq:fit-S}
\end{eqnarray}
This shows that one can consider the grid points as well as the
weights as fitting parameters. The result for 5 equidistant points
with $\Delta r=0.7 \, {\rm fm} $ is shown in Fig.~\ref{delta-shell}
(right panel). Of course, the existence of finite experimental errors
helps in decreasing the number of coarse grained grid points.

\begin{figure}[tb]
\epsfysize=9.0cm
\begin{center}
\begin{minipage}[t]{5 cm}
\epsfig{file=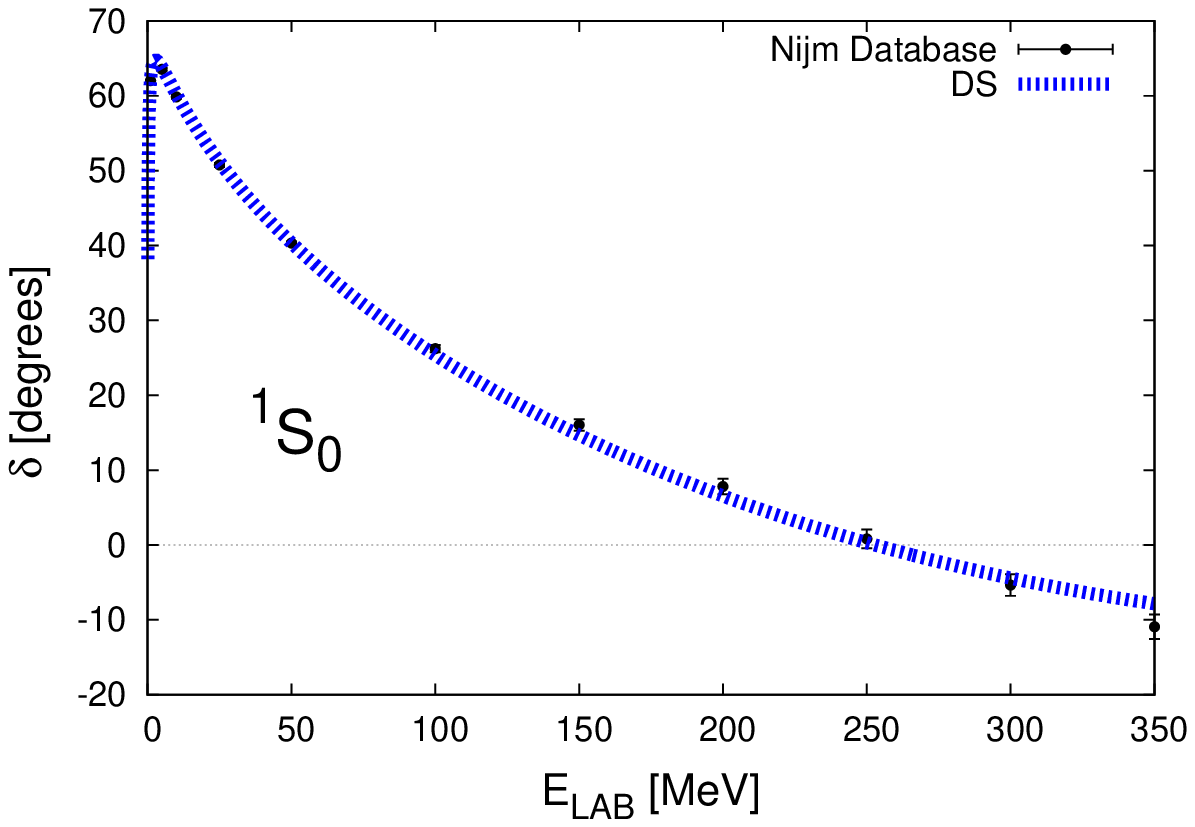,scale=0.4}
\end{minipage}
\begin{minipage}[t]{5 cm}
\epsfig{file=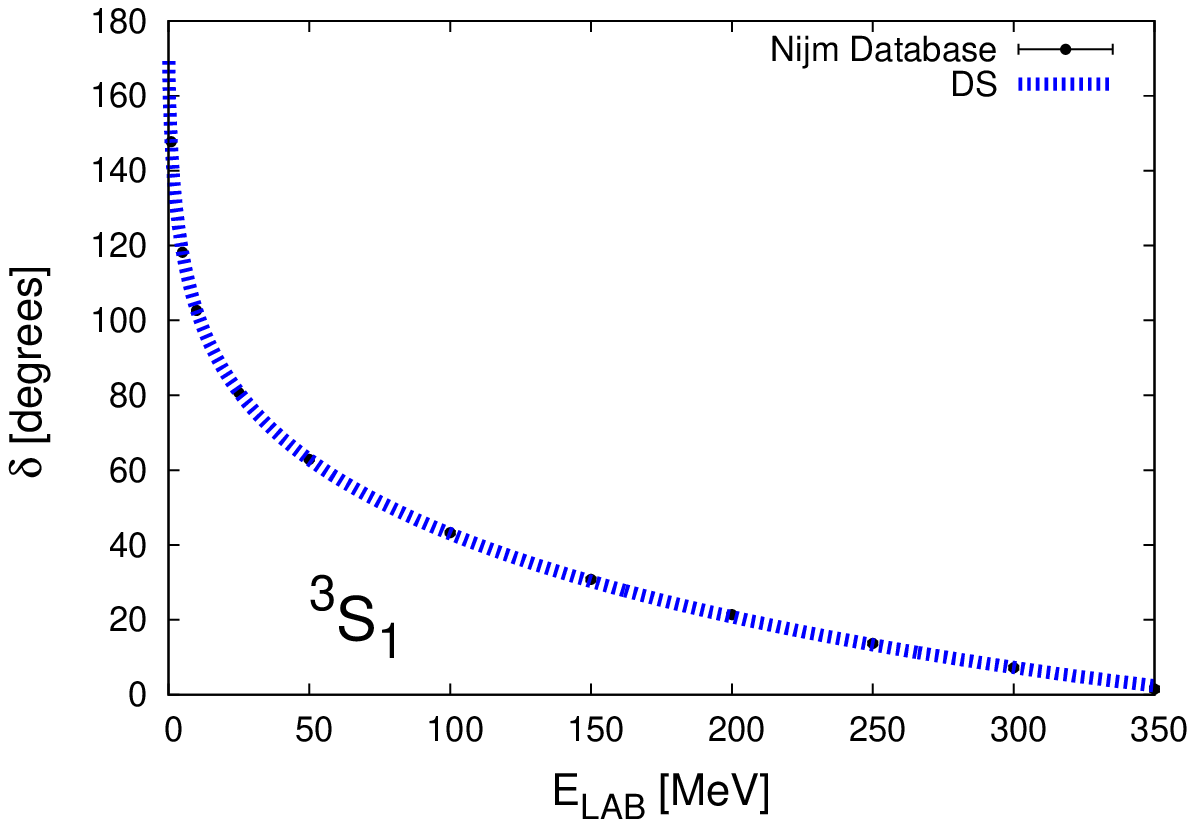,scale=0.4}
\end{minipage}
\begin{minipage}[t]{5 cm}
\epsfig{file=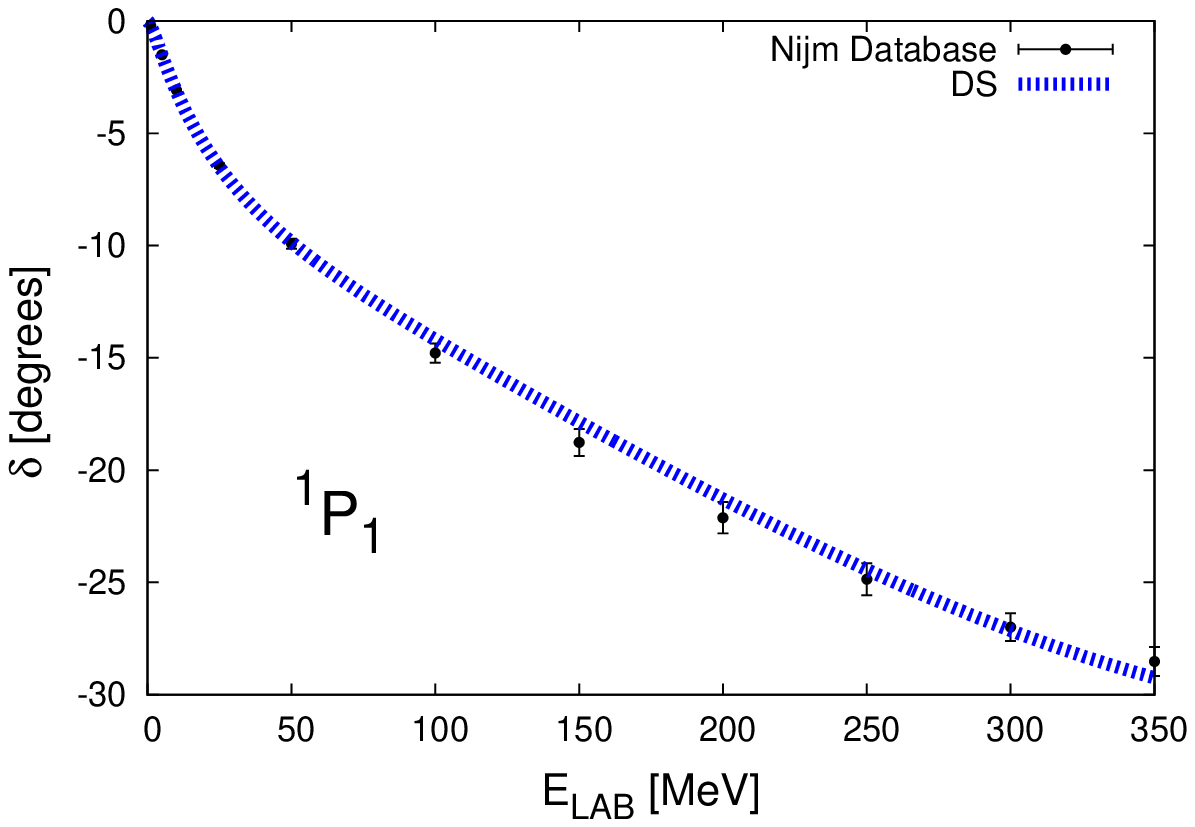,scale=0.4}
\end{minipage}
\begin{minipage}[t]{5 cm}
\epsfig{file=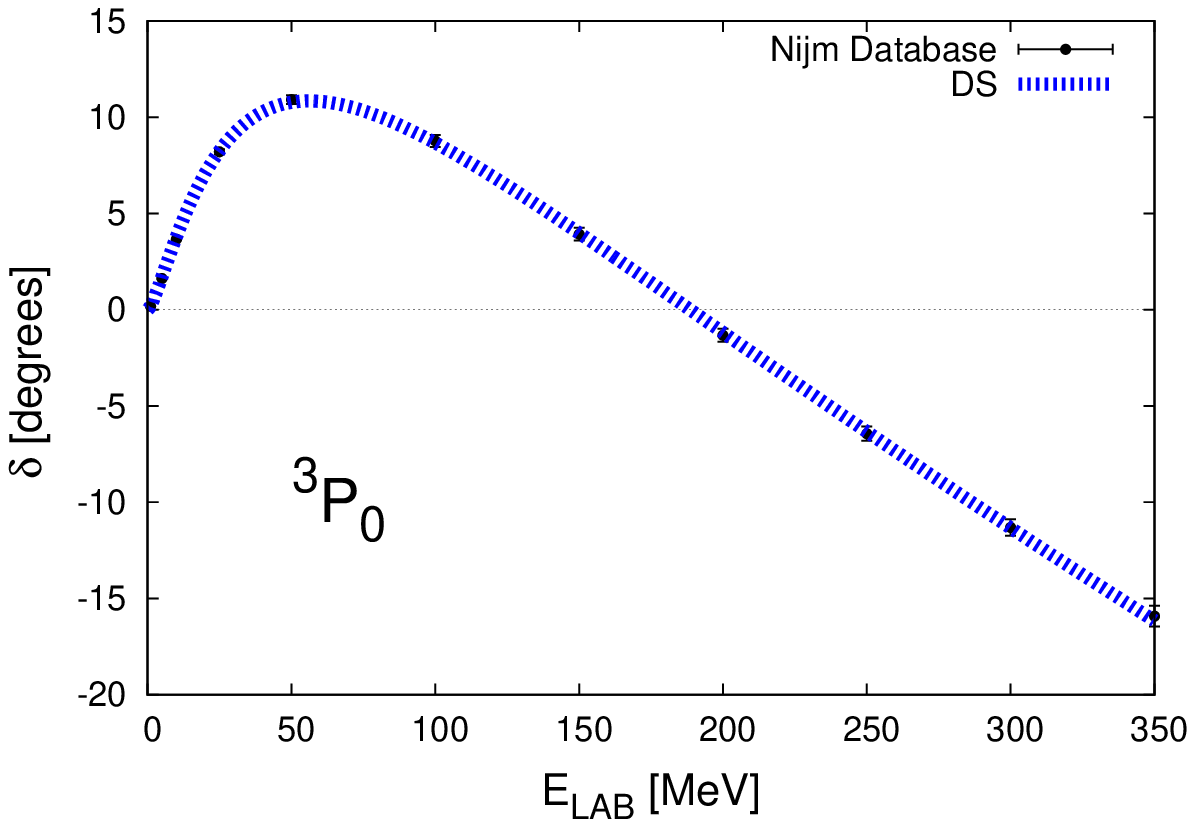,scale=0.4}
\end{minipage}
\begin{minipage}[t]{5 cm}
\epsfig{file=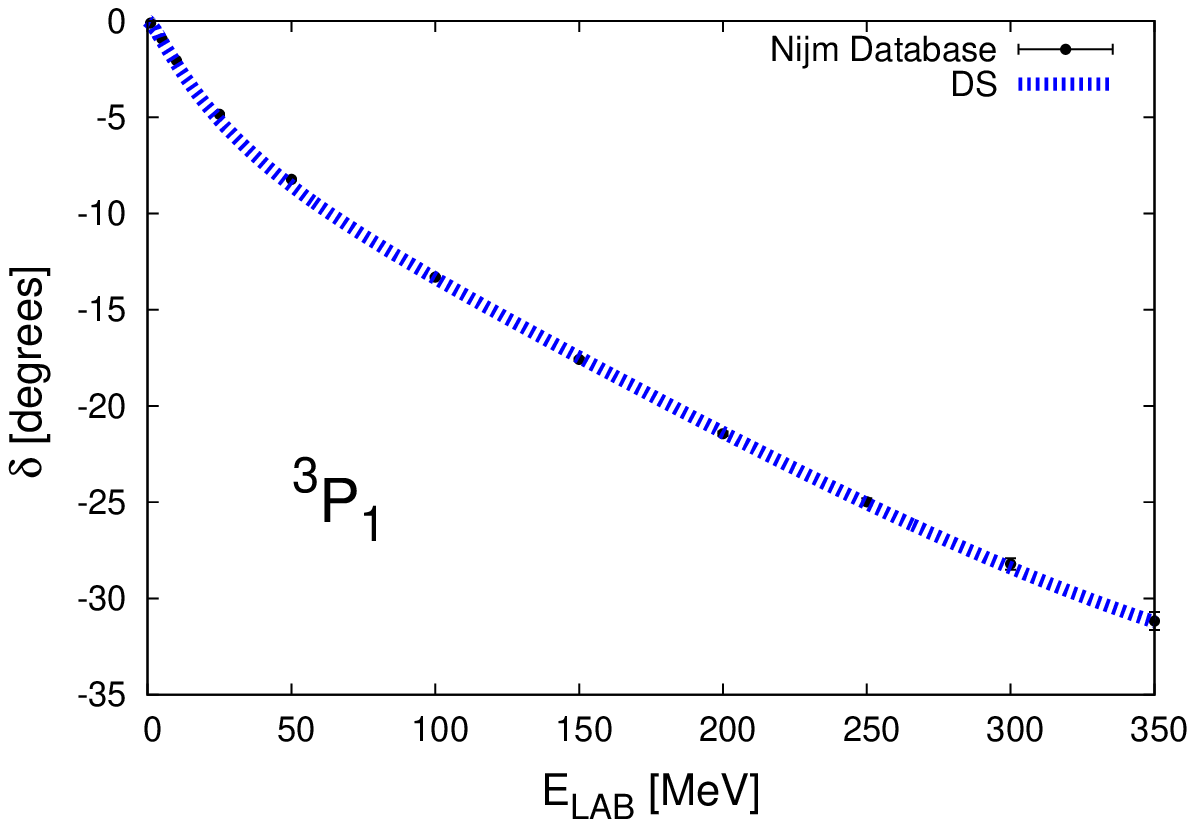,scale=0.4}
\end{minipage}
\begin{minipage}[t]{5 cm}
\epsfig{file=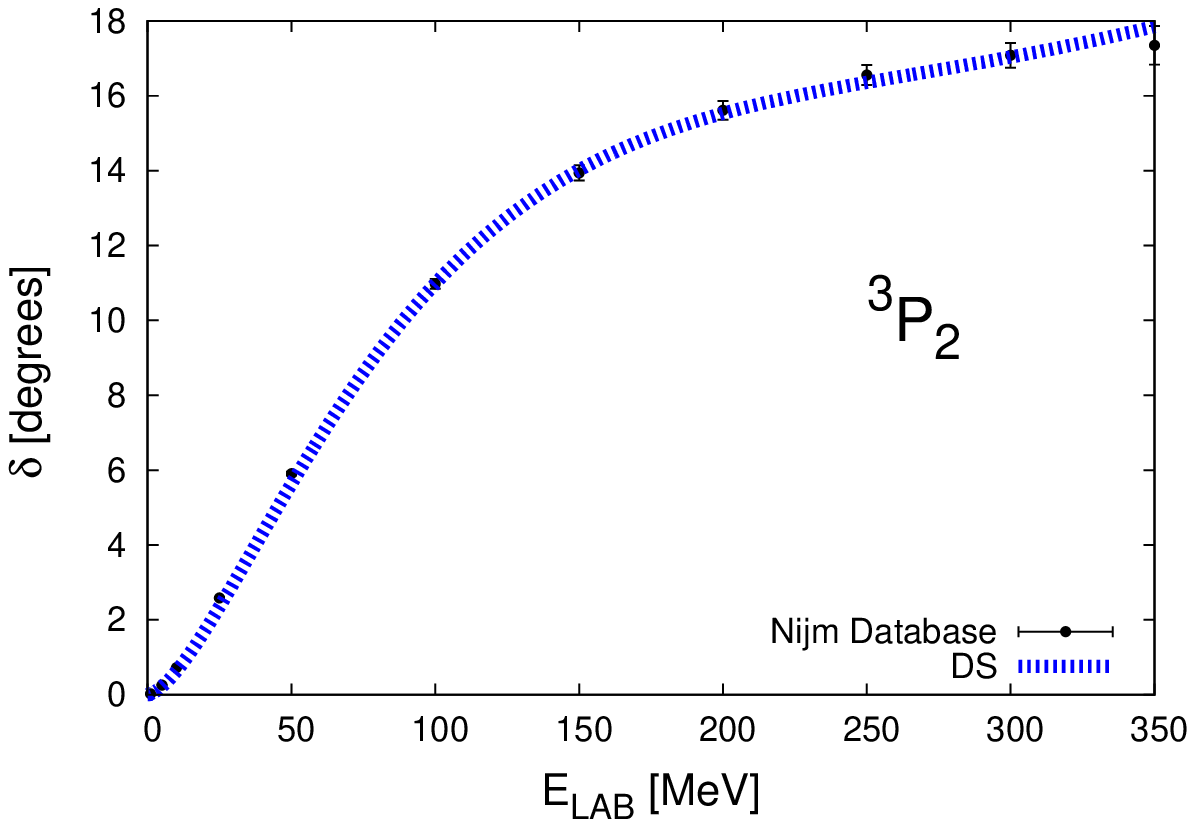,scale=0.4}
\end{minipage}
\begin{minipage}[t]{5 cm}
\epsfig{file=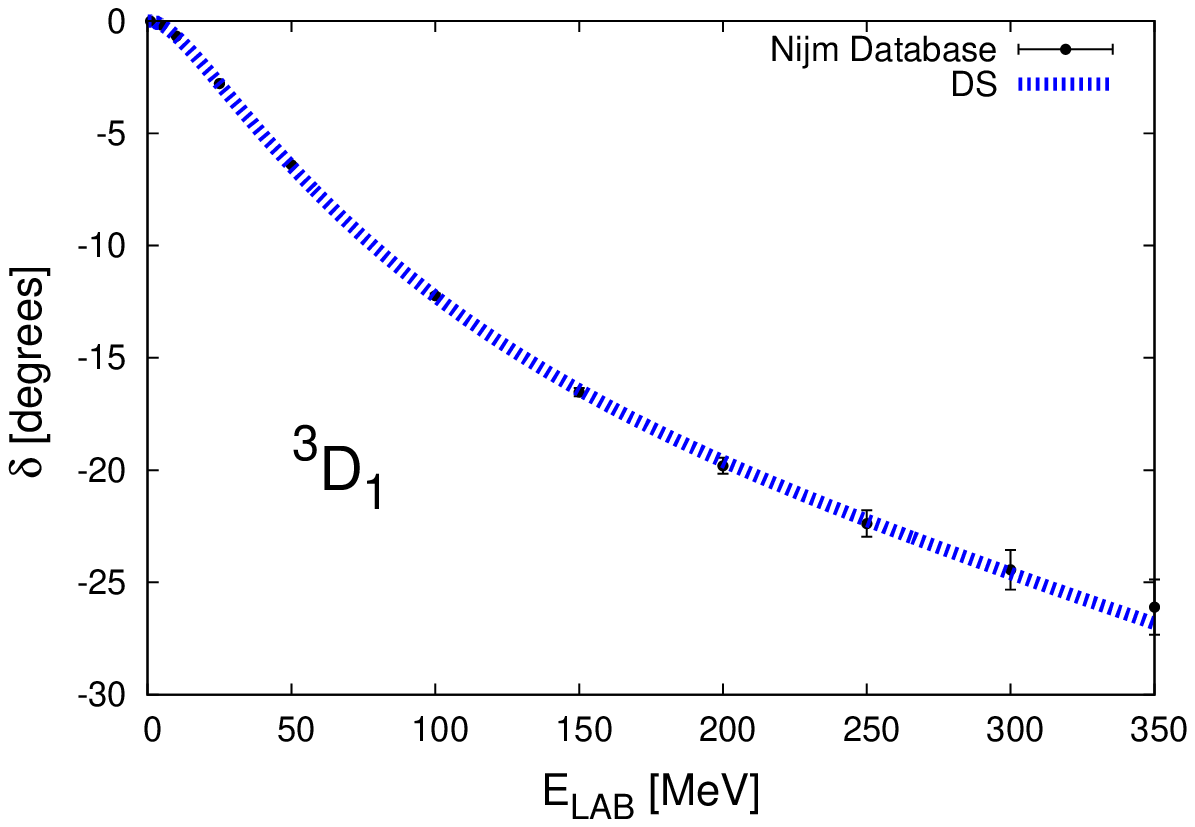,scale=0.4}
\end{minipage}
\begin{minipage}[t]{5 cm}
\epsfig{file=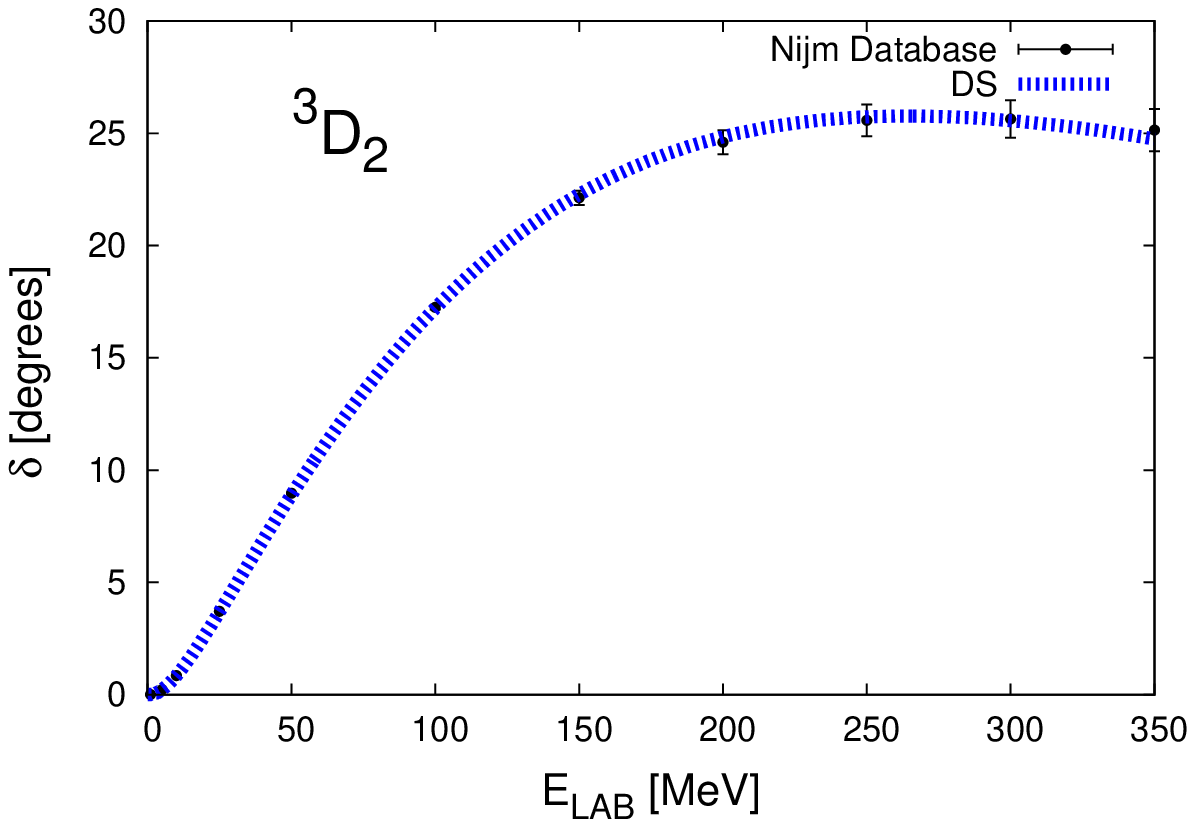,scale=0.4}
\end{minipage}
\begin{minipage}[t]{5 cm}
\epsfig{file=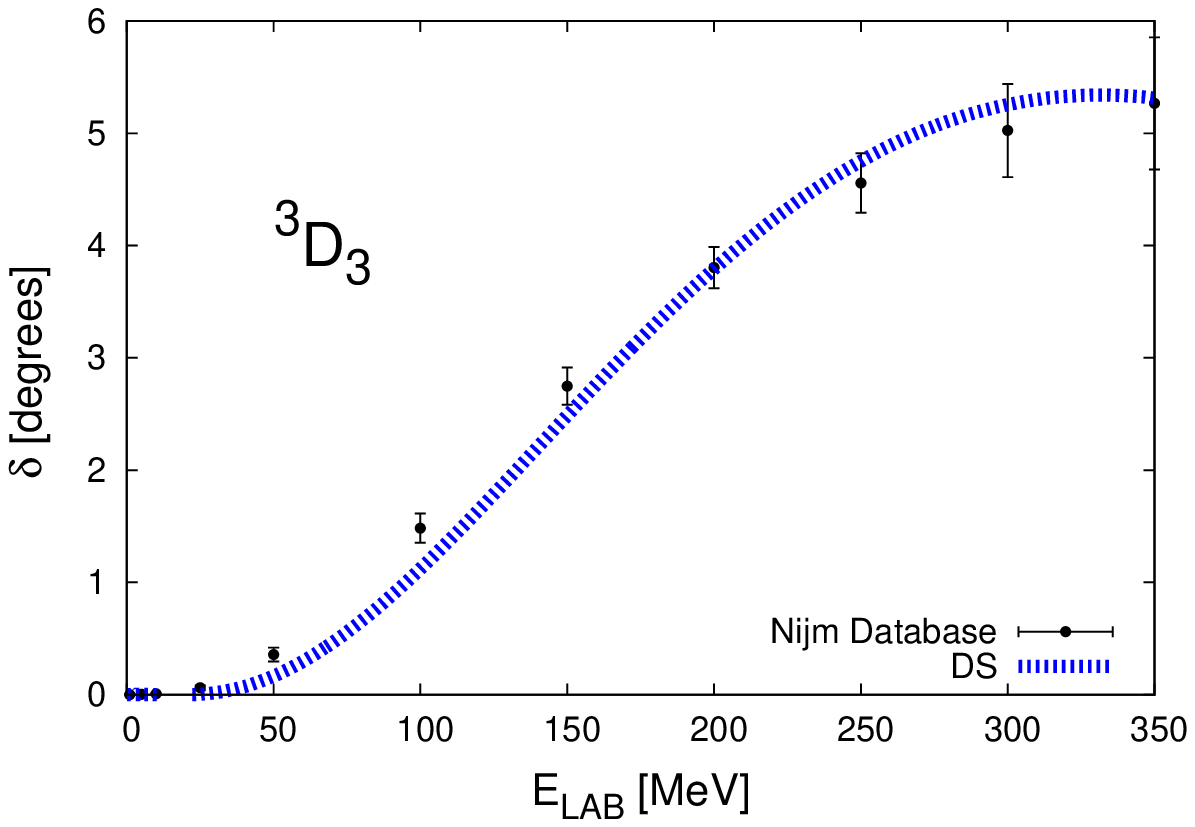,scale=0.4}
\end{minipage}
\begin{minipage}[t]{16.5 cm}
\caption{Phase-shifts fitted to the PWA of Nijmegen in the lowest
  partial waves. \label{fig:phase-shifts}}
\end{minipage}
\end{center}
\end{figure}

We have carried out preliminary fits to the NN
database~\cite{Stoks:1993tb} 
with a pion tail with an average
$m_\pi=138 {\rm MeV}$ starting at $2 {\rm fm}$ for partial waves with
$J \le 4 $ and about 40 parameters with $\chi^2 /{\rm dof} \lesssim
1-2$. The result for low partial waves 
is shown in
Fig.~\ref{fig:phase-shifts}. The full PWA using a CD-OPE potential
tail with the pertinent electromagnetic corrections to the PWA
database will be presented elsewhere.

\begin{figure}[tb]
\begin{center}
\begin{minipage}[t]{8 cm}
\epsfig{file=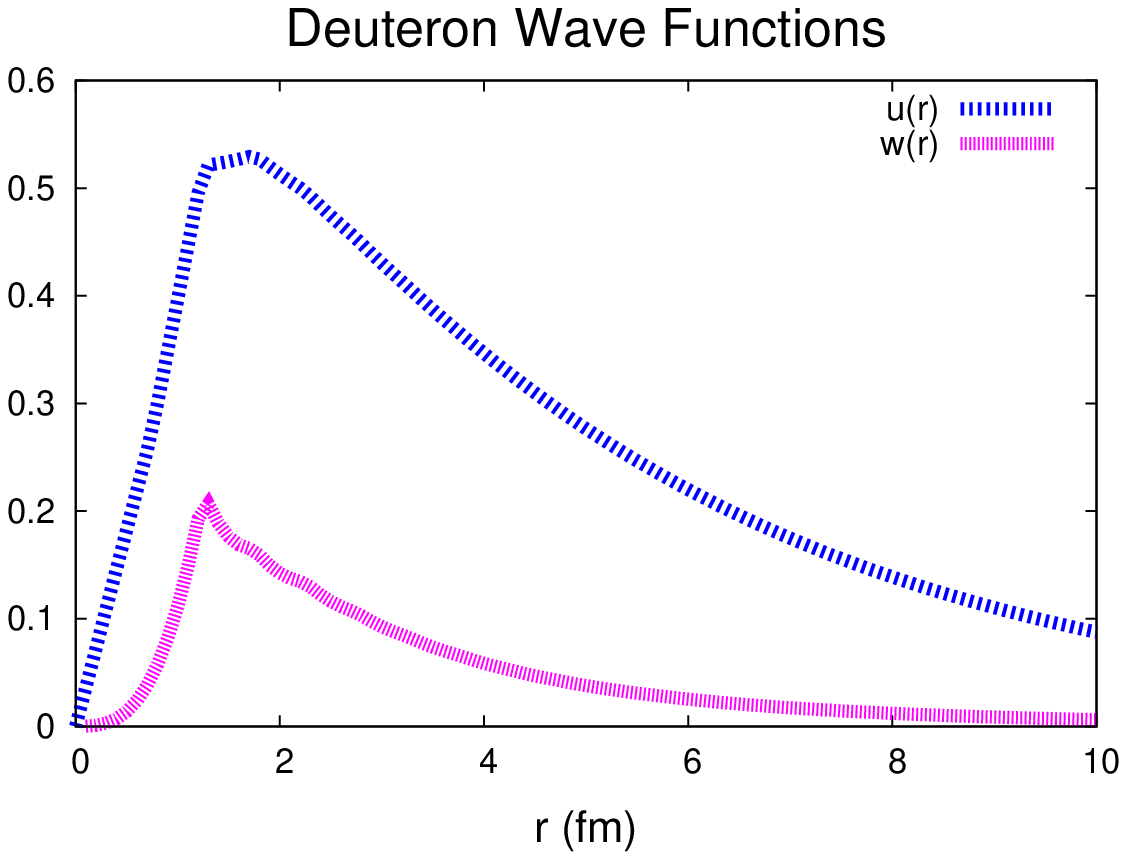,scale=0.5}
\end{minipage}
\begin{minipage}[t]{8 cm}
\epsfig{file=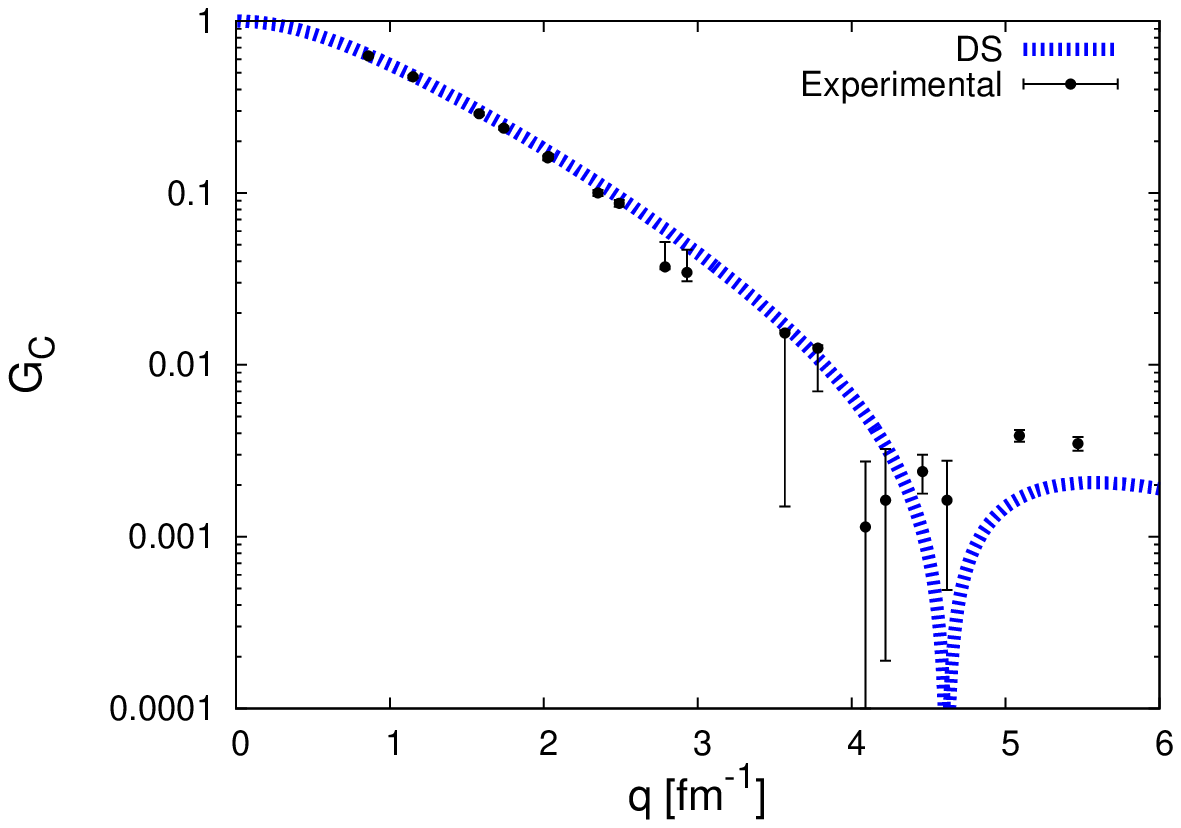,scale=0.5}
\end{minipage}
\begin{minipage}[t]{16.5 cm}
\caption{Left panel: The deuteron wave functions 
Right panel: Charge form factor.\label{fig:deuteron}}
\end{minipage}
\end{center}
\end{figure}


It is straightforward to look at the deuteron by analysing the
$^3S_1-^3D_1$ channel for negative energy. The results can be seen in
Table~\ref{tab:deuteron}. The deuteron wave functions as well as the
corresponding charge form factor is displayed in
Fig.~\ref{fig:deuteron}. The peaks in the wave functions correspond to
the discontinuity in the derivatives at the chosen grid points which,
as wee can see, does not become dramatic for the form factor at the
considered $q$'s.

\begin{table}
\begin{center}
\begin{minipage}[t]{16.5 cm}
\caption{Deuteron parameters}
\label{tab:deuteron}
\end{minipage}
\begin{tabular}{|c|l|l|l|l|l|l|} 
\hline
& Delta Shell & Empirical & Nijm I~\cite{Stoks:1994wp}   & Nijm II~\cite{Stoks:1994wp}  & Reid93~\cite{Stoks:1994wp}   & AV18~\cite{Wiringa:1994wb}  \\
      \hline
		$\gamma ({\rm fm}^{-1})$ & 0.230348 & 0.231605       & Input    & Input    & Input    & Input             \\
		$\eta$   & 0.02488  & 0.0256(5)      & 0.02534  & 0.02521  & 0.02514  & 0.0250             \\
		$A_S ({\rm fm}^{1/2})$ & 0.8768   & 0.8781(44)     & 0.8841   & 0.8845   & 0.8853   & 0.8850      \\
		$r_m ({\rm fm})$    & 1.9676   & 1.953(3)       & 1.9666   & 1.9675   & 1.9686   & 1.967              \\
		$Q_D ({\rm fm}^{2}) $  & 0.2693   & 0.2859(3)      & 0.2719   & 0.2707   & 0.2703   & 0.270     \\
		$P_D$    & 5.498    & 5.67(4)        & 5.664    & 5.635    & 5.699    & 5.76               \\
\hline
	\end{tabular} 
\end{center}
\end{table}


Fourier transforming the DS potential in the S-waves gives
 \begin{equation}
  \langle k' | V^{SJ}_{l' l} | k \rangle = \sum_i (\lambda_i)^{JS}_{l',l} r_i^2 j_{l'} (k' r_i) j_l (k r_i) \, , 
\label{eq:V_lowk}
 \end{equation}
which is a finite rank separable potential, a representation which
proved very handy in the past for few-body and nuclear matter
calculations (see e.g.  \cite{Grygorov:2010re}). 

\begin{figure}[tb]
\begin{center}
\begin{minipage}[t]{8 cm}
\epsfig{file=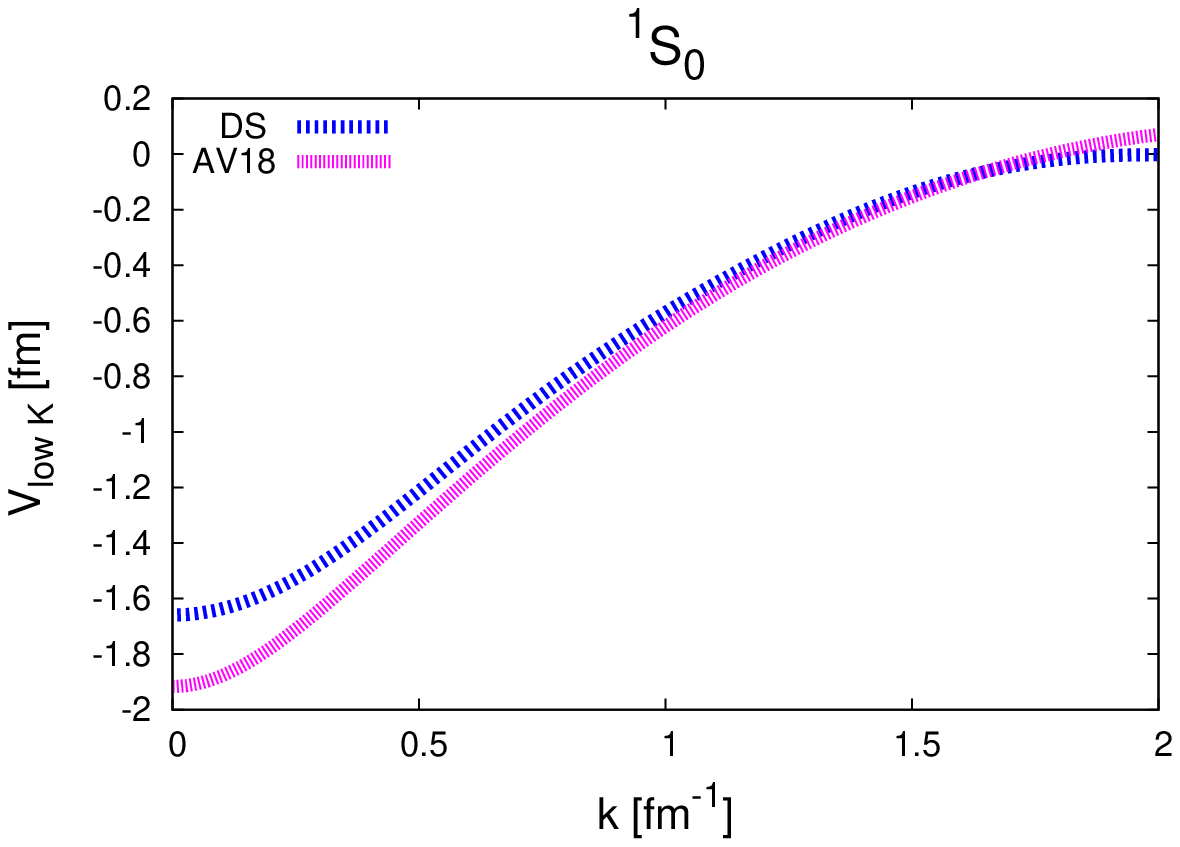,scale=0.5}
\end{minipage}
\begin{minipage}[t]{8 cm}
\epsfig{file=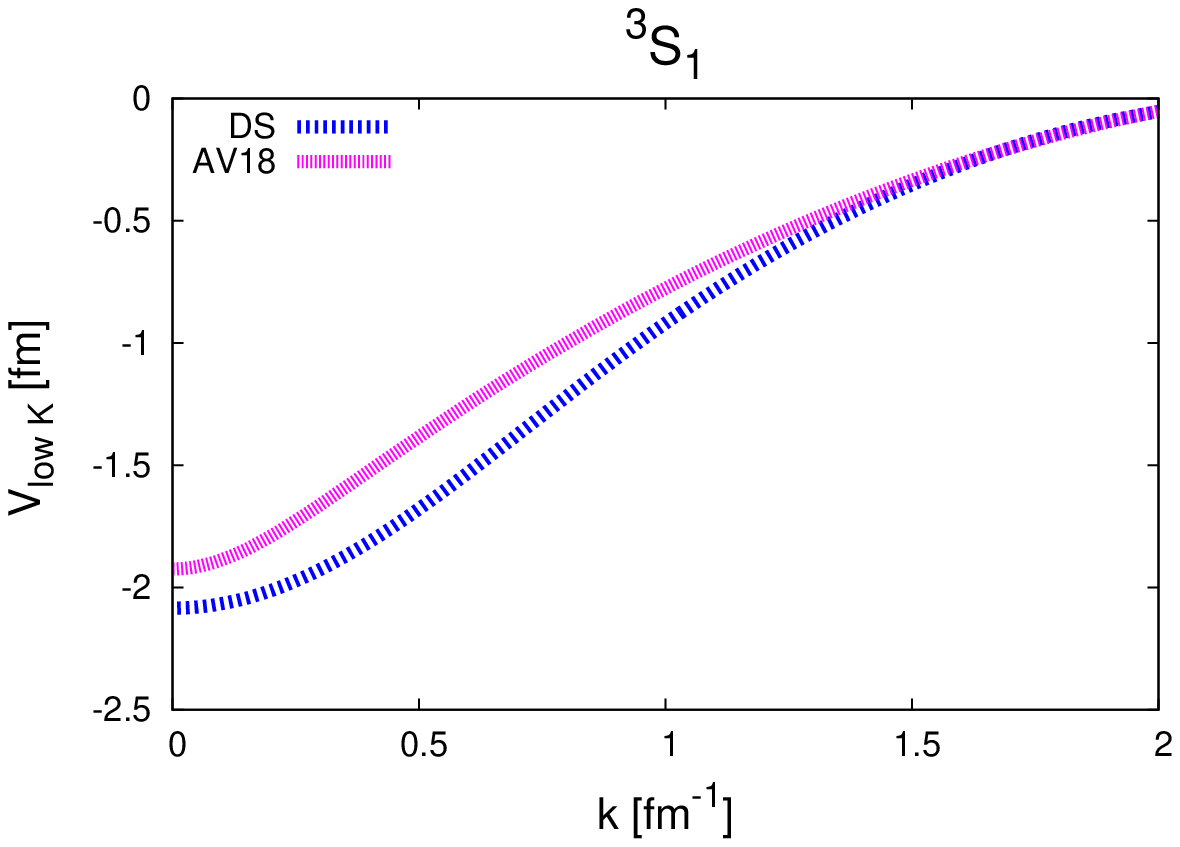,scale=0.5}
\end{minipage}
\begin{minipage}[t]{16.5 cm}
\caption{Comparison of the coordinate space coarse-grained potentials
  in the $^1S_0$ and $^3S_1$ channels with the corresponding $V_{\rm
    low k}$ potentials~\cite{Bogner:2003wn}, obtained from the
  AV18~\cite{Wiringa:1994wb} interaction.\label{fig:V_lowk}}
\end{minipage}
\end{center}
\end{figure}

We show in Fig.~\ref{fig:V_lowk} a comparison of the Fourier
transformed DS potentials in the $^1S_0$ and $^3S_1$ channels with
with parameters as in Eq.~(\ref{eq:fit-S}) to the corresponding
diagonal elements of the $V_{\rm low k}$
potentials~\cite{Bogner:2003wn}, obtained from the
AV18~\cite{Wiringa:1994wb} interaction.  While the resemblance is
indeed rather close, we do not expect a perfect description since the
way the scattering problem is treated in the DS case is {\it
  different} as in the $V_{\rm low_k}$ case. We have checked that one
can represent quite accurately the current diagonal pieces of the
$V_{\rm low k}$ potentials~\cite{Bogner:2003wn} by
Eq.~(\ref{eq:V_lowk}), but does not necessarily reproduce the
off-diagonal matrix elements constructed in the $V_{\rm low k}$
approach from the truncated the half-off shell Lippmann-Schwinger
equation by a specific block-transformation method.

\begin{figure}[tb]
\epsfysize=9.0cm
\begin{center}
\begin{minipage}[t]{5.5 cm}
\epsfig{file=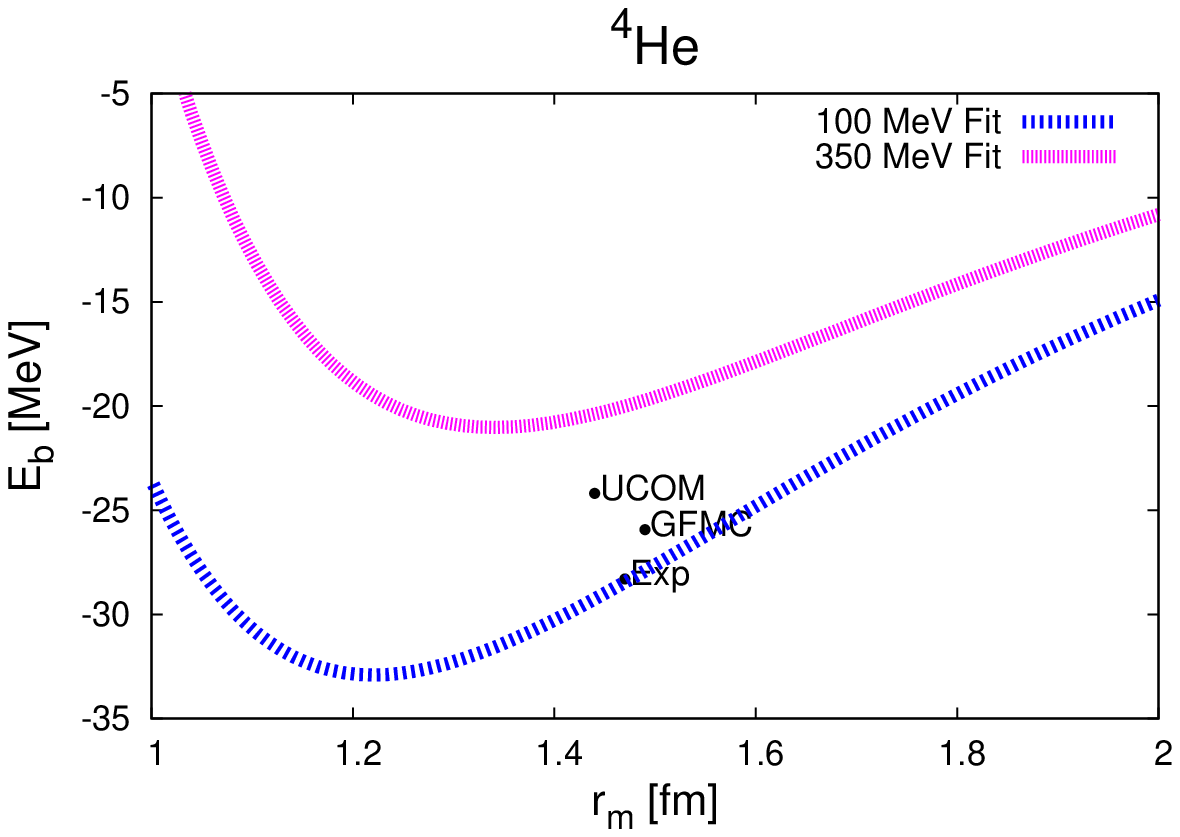,scale=0.45}
\end{minipage}
\begin{minipage}[t]{5.5 cm}
\epsfig{file=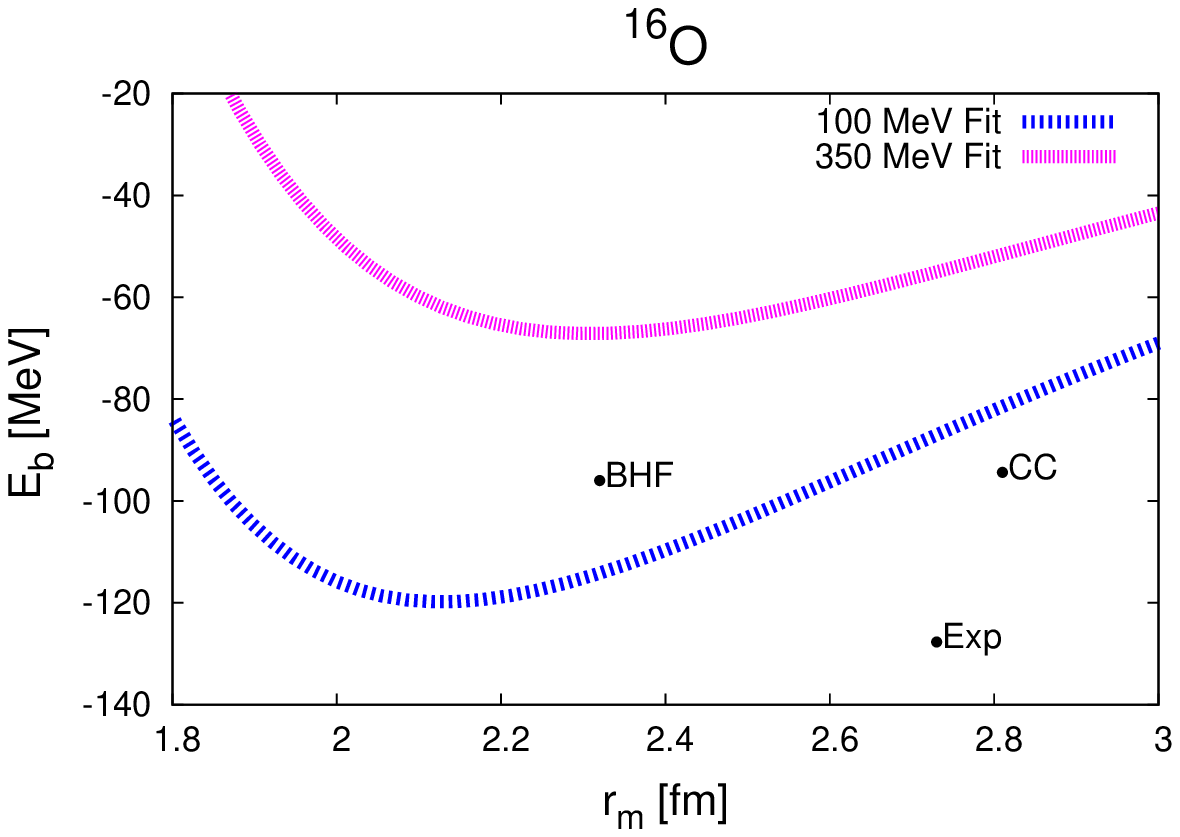,scale=0.45}
\end{minipage}
\begin{minipage}[t]{5.5 cm}
\epsfig{file=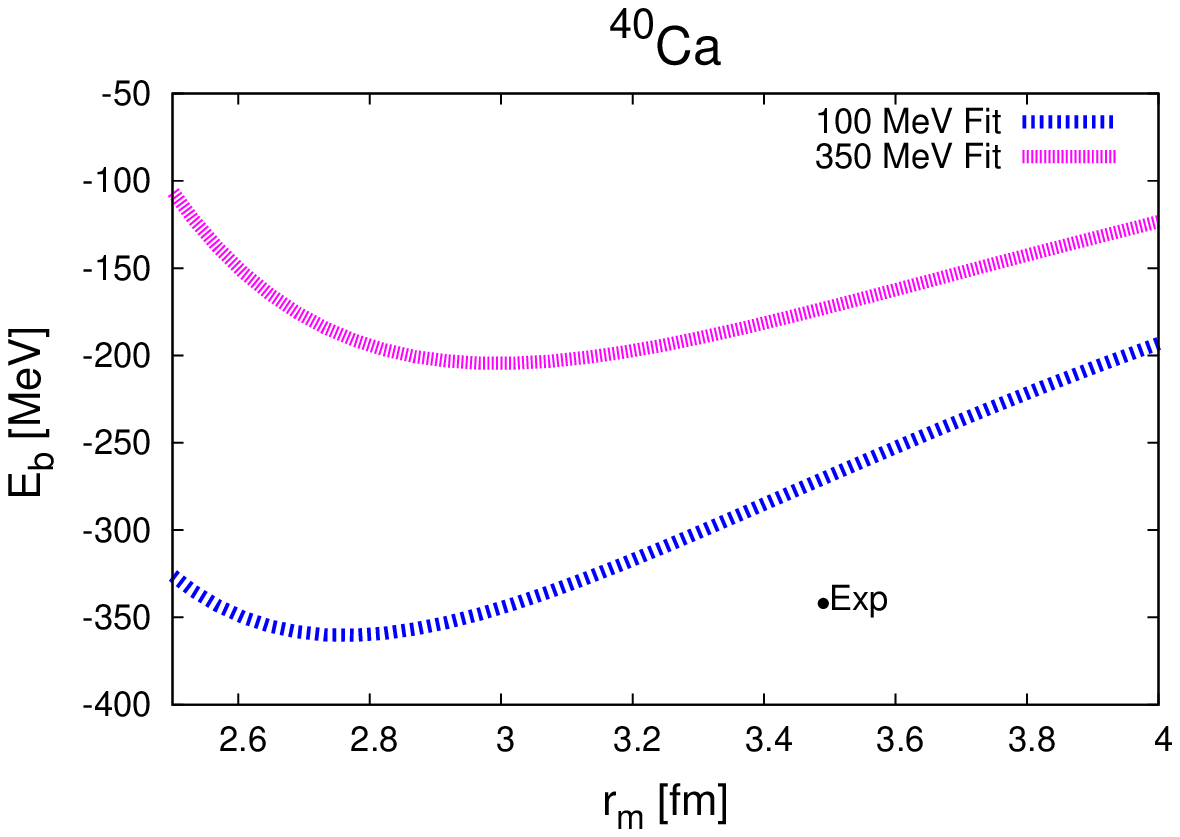,scale=0.45}
\end{minipage}
\begin{minipage}[t]{16.5 cm}
\caption{Binding energy of closed-shell nuclei $^4{\rm He}$,
  $^{16}{\rm O}$ and $^{40}{\rm Ca}$ using a HO shell model Slater
  determinant as a function of the msr when the relevant phase shifts
  are reproduced up to $E_{\rm LAB}= 100 {\rm MeV}$ and $E_{\rm LAB}=
  350 {\rm MeV}$.  We compare to other methods (see main text).
\label{fig:binding}}
\end{minipage}
\end{center}
\end{figure}

\section{Closed-shell nuclei}

When switching from the NN problem to the many body nuclear problem
the features and the form of the interaction are relevant in terms of
computational cost and feasibility. We coarse-grain the interaction,
but keep the exact kinetic energy, so that 
for two nucleons at a relative
distance  $| \vec x_1 - \vec x_2| \neq r_c$ the
interaction vanishes, and hence the wave function becomes a Slater
determinant of single particle states
\begin{eqnarray}
\psi (\vec x_1, \dots, \vec x_A) = {\cal A } \left[ \phi_{n_1,l_1, s,
    m_{s1}, t, m_{t_1}} ( \vec x_1) \dots \phi_{n_A,l_A, s, m_{sA}, t,
    m_{t_A}} (\vec x_A) \right] \, . 
\end{eqnarray}
We use Harmonic Oscillator single particle wave functions with
oscillator parameter $b$, where the spurious CM motion is 
exactly subtracted,  for the shell-configurations
$^4$He:$(1s)^4$, $^{16}$O:$(1s)^4 (1p)^{12}$ and $^{40}$Ca:$(1s)^4
(1p)^{12} (2s)^4 (1d)^{20}$. Generally, for double-closed shell nuclei
one has 
\begin{eqnarray}
\langle V_2 \rangle_A = \sum_{nl JS} g_{nl JS} \langle nl | V^{JST} |
nl \rangle \, , 
\end{eqnarray}
in terms of the relative matrix elements and $g_{nlsj}$ depends on the
Talmi-Moshinsky brackets~\footnote{For instance, for $^4{\rm He}$ one
  has a m.s.r. $r_m = 3 b/2 \sqrt{2}$ and for $R_{1s}(r)=2
  e^{-\frac{r^2}{2 b^2}}/\sqrt[4]{\pi } b^{3/2} $,
$$
\langle T \rangle_{^4{\rm He}} + \langle V_2 \rangle_{^4{\rm He}} 
= 3 \langle 1s |\frac{p^2}{2M} | 1s \rangle  + 6 \, \langle 1s | \frac12
\left(V_{^1S_0} + V_{^3S_1} \right) | 1s \rangle =  \frac{9}{4b^2M} + 
\frac{6}{b^3 M} \sqrt{\frac{2}{\pi }}  \sum_n 
( \lambda_{n, ^1S_0} + \lambda_{n, ^3S_1} )
r_n^2 e^{-\frac{r_n^2}{2 b^2}}
$$ For an Android implementation of these calculations see e.g.
\url{http://www.ugr.es/~amaro/android/})}. Using the single delta
function which is just fixed with the S-waves scattering lengths and
effective ranges (see below Eq.~(\ref{eq:ere}) one obtains at the
minimum $B ( ^4{\rm He})= 20 {\rm MeV}$. In common with other soft
potentials~\cite{Afnan:1968zj} the interaction does not require strong
correlations in the many-body wave function. This is due to the fact
that since the phase-shift is reproduced to about $E_{\rm LAB}= 50
{\rm MeV}$ the core may be ignored. Clearly, if we insist on
reproducing up to $E_{\rm LAB}= 350 {\rm MeV}$ a strongly repulsive DS
contribution emerges and thus a product wave function is not
appropriate. One can improve on this by adding more deltas as in
Eq.~(\ref{eq:fit-S}) but keeping the fit to $E_{\rm LAB} \le 100 \,
{\rm MeV}$, in which case $B ( ^4{\rm He})= 24 {\rm MeV}$. This is
surprisingly close to the Green Function Monte Carlo (GFMC) AV18
\cite{Pieper:2001mp} and the UCOM method~\cite{Neff:2002nu} without
three-body forces and complies to a cancellation between the core in the and
the correlations the wave function. 

The results for the binding energy as a function of the corresponding
msr radius are presented in Fig.~\ref{fig:binding}. We compare with
the UCOM method~\cite{Neff:2002nu}, Brueckner-Hartree-Fock
(BHF)~\cite{Muther:2000qx} and Coupled Cluster
(CC)~\cite{Heisenberg:1998qn}.  In the UCOM method~\cite{Neff:2002nu}
a unitary local transformation generates a smooth nonlocal interaction
from the AV18-potential while the  wave functions are the
same. As advertised, our results depend on the fitted energy range,
somewhat resembling analogous ambiguities as those of the UCOM.

\section{Conclusions}

We have shown how sampling of the NN interaction by a delta shell
potential with a resolution determined by the deBroglie wavelength of
the most energetic particle provides a coarse graining in
configuration space, analogous to the $V_{\rm lowk}$
approach. However, rather than transforming a high quality potential
we suggest to determine the NN coarse grained interaction directly
from the scattering data. A preliminary fit to the np phase shifts in
the Nijmegen data base to all partial waves with $j \leq 4$ requires
about 40 fitting parameters yielding $\chi^2/\text{d.o.f.} \lesssim 2$
(less than 1 in some waves). Deuteron properties show good agreement
with empirical values and other calculations. Harmonic oscillator
shell model variational calculations of nuclear binding energies
provide results at the 20-30$\%$ accuracy.


The work is supported by Spanish DGI and FEDER funds (grant FIS2008-
01143/FIS) and Junta de Andaluc{\'{\i}a} (grant FQM225).  R.N.P. is
supported by a Mexican CONACYT grant.

\end{document}